\documentclass[runningheads]{llncs}

\usepackage[T1]{fontenc}

\usepackage[ruled,vlined,linesnumbered]{algorithm2e}
\usepackage{hyperref}
\usepackage{amsmath,amssymb}
\usepackage{pgfplots}
\usepackage{graphicx}
\usepackage{caption}
\usepackage{subcaption}
\usepackage{mathtools}
\usepackage{multirow}
\usepackage{booktabs}
\usepackage{enumitem}

\allowdisplaybreaks

\newcommand{\state}{s}

\newcommand{\action}{a}
\newcommand{\maction}{\hat{a}}
\newcommand{\actions}{A}

\newcommand{\policy}{\pi}

\newcommand{\reals}{\mathbb{R}}

\newcommand{\gaussian}{\mathcal{N}}
\newcommand{\State}{s}

\newcommand{\States}{S}
\newcommand{\Actions}{A}

\newcommand{\Transitions}{T}
\newcommand{\Action}{a}
\newcommand{\model}{M}
\newcommand{\param}{\mathbf{\theta}}

\newcommand{\timeid}{t}
\newcommand{\ellipsoid}{\mathcal{E}}
\newcommand{\ellipsoidbound}{\mathcal{H}}
\newcommand{\rr}{\mathcal{R}}
\newcommand{\adaptedaction}{\hat{\mu}_{\action_t} }
\DeclareMathOperator*{\argmin}{arg\,min}

\newcommand{\myipara}[1]{\vspace{0.5em} \noindent{\em #1}.} 

\newcommand{\randvar}{r}
\newcommand{\avg}[1]{\mu_{#1}}
\newcommand{\expected}{\mathop{\mathbb{E}}}

\begin{document}

\title{Convex Optimization-based Policy Adaptation to Compensate for Distributional Shifts}% If the paper title is too long for the running head, you can set
% an abbreviated paper title here
%
\titlerunning{Convex Optimization-based Policy Adaptation }

\author{Navid Hashemi\inst{1},  Justin Ruths \inst{2} \and  Jyotirmoy Deshmukh \inst{1} }
% %
\authorrunning{N. Hashemi et al.}
% % First names are abbreviated in the running head.
% % If there are more than two authors, 'et al.' is used.
% %
\institute{University of Southern California, Los Angeles CA, USA \and
University of Texas at Dallas, Dallas TX, USA }
% %

\maketitle

\begin{abstract}

Many real-world systems often involve physical components or operating
environments with highly nonlinear and uncertain dynamics. A number of
different control algorithms can be used to design optimal controllers
for such systems, assuming a reasonably high-fidelity model of the actual
system. However, the assumptions made on the stochastic dynamics of the
model when designing the optimal controller may no longer be valid when
the system is deployed in the real-world. The problem addressed by
this paper is the following: Suppose we obtain an optimal trajectory 
by solving a control problem in the training environment,
how do we ensure that the real-world system trajectory tracks this optimal trajectory
with minimal amount of error in a deployment environment. In other
words, we want to learn how we can adapt an optimal trained policy
to distribution shifts in the environment.
Distribution shifts are problematic in 
safety-critical systems, where a trained policy may lead to unsafe 
outcomes during deployment. 
We show that this problem can be cast as a  nonlinear optimization problem 
that could be solved using heuristic method such as particle swarm optimization 
(PSO). However, if we instead consider a convex relaxation of this problem, we
can learn policies that track the optimal trajectory with much
better error performance, and faster computation times.
We demonstrate the efficacy of our approach on
tracking an optimal path using a Dubin's car model, 
and collision avoidance using
both a linear and nonlinear model for adaptive cruise control.

% Many real-world systems often involve physical components or operating environments with highly nonlinear and uncertain dynamics. A number of different control algorithms can be used to design optimal controllers for such systems, assuming a reasonably high-fidelity model of the actual system. However, the assumptions made on the stochastic dynamics of the model when designing the optimal controller may no longer be valid when the system is deployed in the real-world. The problem addressed by this paper is the following: Suppose we obtain an optimal trajectory by solving a control problem in the training environment, how do we ensure that the real-world system trajectory tracks this optimal trajectory with minimal amount of error in a deployment environment. In other words, we want to learn how we can adapt an optimal trained policy
% to distribution shifts in the environment. Distribution shifts are problematic in safety-critical systems, where a trained policy may lead to unsafe outcomes during deployment. 
% We show that this problem can be cast as a  nonlinear optimization problem  that could be solved using heuristic method such as particle swarm optimization  (PSO). However, if we instead consider a convex relaxation of this problem, we
% can learn policies that track the optimal trajectory with much
% better error performance, and faster computation times. We demonstrate the efficacy of our approach on tracking an optimal path using a Dubin's car model,  and collision avoidance using both a linear and nonlinear model for adaptive cruise control.

\end{abstract}

\section{Introduction}
% Data-driven control-design methods have become increasingly 
% popular  in settings where the symbolic expressions for the
% dynamics of the  plant model are difficult to obtain, and 
% where the operating environment is highly uncertain. Some 
% examples of such techniques include neural predictive control
% \cite{ }, data-driven predictive control \cite{ }, and
% deep model-based reinforcement learning \cite{ }. These methods
% all learn high-fidelity data-driven models of the plant/
% environment, and can give probabilistic safety guarantees of 
% the learned controller under the assumption that the learned
% model is a close representation of the actual system dynamics.

Systems operating in highly uncertain environments are often
modeled as stochastic dynamical systems that satisfy Markov
assumptions, i.e. Markov decision processes (MDPs). 
Given a state $s_t$ (i.e., the state at time $t$), a (discrete-time) 
MDP defines a distribution on $s_{t+1}$
conditioned on $s_t$ and the control action at time $t$ (denoted $a_t$). We
call this distribution the transition dynamics.
For such systems, a number of model-based and
data-driven control design methods have been explored to learn
an optimal policy (i.e. a function from the set of states to
the set of actions) that minimizes some 
control-theoretic cost function \cite{khalil1996robust}. Model-based methods 
explicitly, and data-driven methods implicitly, assume a specific
distribution for the transition dynamics. However, when the system
is deployed in the real-world, this distribution may not be the 
same; this change in distribution is called a {\em distribution
shift}.

The fundamental problem addressed by this paper is adapting a
pre-learned control policy to compensate for distribution shifts. While
it is possible to retrain the control policy on the new environment, 
it is typically expensive to learn the optimal control policy. However,
a crucial observation that we make is that while learning an optimal
policy is expensive, learning a reasonable high-fidelity model of 
the transition dynamics may be feasible. We call such a learned
model a {\em surrogate model}.
In this paper, we show that under
certain kinds of distribution shifts, the problem of
adapting an existing optimal policy to the new deployment environment
can be framed as a nonlinear optimization problem over the optimal
trained trajectory and the surrogate model. Furthermore, we show
that if the surrogate is a neural network (with rectified linear
unit or ReLU based activation), then there is a convex relaxation
of the original optimization problem. This convex relaxation
permits an efficient procedure to find a modified action that
minimizes the error between the optimal trained trajectory and the 
system trajectory in the deployment environment.
Finally, we empirically show that if the trained trajectory meets 
desired objectives of safety, then such policy adaptation can
provide safety in the deployment setting.

The main technical idea in our work is inspired by recent work in 
\cite{fazlyab2019safety}, where
the authors proposed an efficient method to provide probabilistic bounds 
on the output of a neural network, given a Gaussian distribution on its 
inputs. We show how we can use this result to propagate the effects of a 
distribution shift. However, the result in \cite{fazlyab2019safety} does
not consider the problem of finding optimal actions (which is a non-convex
problem). In this research, we propose a methodology to convexify this result for finding optimal actions. 
We demonstrate our 
technique on a tracking problem using a Dubin's car model and a
collision avoidance problem that uses adaptive cruise control.

The rest of the paper is organized as follows. In Section~\ref{sec:prelim} we discuss the preliminaries, terminology and technical notation. In Section~\ref{sec:adaptation}, we discuss our policy adaptation approach, and provide experimental results in Section~\ref{sec:results}. We conclude with related work in Section~\ref{sec:conclusion}.

\section{Preliminaries}
\label{sec:prelim}

\myipara{Notation} A multi-variate Gaussian distribution
is denoted as $\gaussian(\mu, \Sigma)$, where $\mu$ and $\Sigma$ 
represent the mean vector and covariance, respectively. For a 
Gaussian-distributed random vector $\randvar \in \reals^n$, we
denote its mean value by $\avg{r}$ and its covariance by 
$\Sigma_r$.  Let $c \in \reals^n$, then an ellipsoid centered 
at $c$ with the {\em shape matrix} $\Omega$ is denoted as 
$\ellipsoid(c,\Omega)$, 
i.e., $\ellipsoid(c,\Omega) = \left\{ x\, \middle| (x-c)^\top 
\Omega^{-1} (x-c) \leq 1 \right\}$.  Given a non-convex set, 
$\mathcal{Y}$  we use the notation $\ellipsoidbound(\mathcal{Y})$ to 
denote the set of ellipsoids that contain $\mathcal{Y}$, i.e.,
$\left\{ \ellipsoid(c,\Omega)\, \middle|\,\mathcal{Y} \subseteq 
\ellipsoid(c,\Omega)  \right\}$.

\myipara{Markov Decision Process, Optimal Policy} We now formalize the
notion of the type of stochastic dynamical systems that we address
in this paper as a Markov Decision process.
\begin{definition}[Markov Decision Process (MDP)] 
A Markov decision process is a tuple $\model = (\States,\Actions,
\Transitions, \iota)$, where $\States$ and $\Actions$ denote the 
set of states  and actions respectively, $\Transitions(\State'\mid \State,
\Action)$ is the probability distribution on the next state conditioned
on the current state and action, and $\iota$ is a distribution on $\States$
that is sampled to identify an initial state of the MDP\footnote{Technically, 
this definition pertains to the transition structure of a stochastic dynamical 
system. Typically, dynamical systems are defined in terms of difference or 
differential equations describing the temporal evolution of a state variable. 
We assume that $\Transitions(\State'|\State,\action)$ is thus the infinite set
of transitions consistent with any given system dynamics.}. 
\end{definition}

In our approach, we are interested in {\em finite}-{\em horizon}
trajectories sampled from the MDP's transition dynamics. A {\em policy} 
$\policy(\action\mid\State)$ of the MDP is a distribution on the set of actions conditioned on the current state. Given a fixed
policy of the MDP, a $T$-length {\em trajectory} (denoted $\tau$) or 
behavior of the MDP is a  sequence of states $\State_0,\ldots, \State_T$ such 
that $\State_0 \sim \iota$, and for all $t \in [0,T-1]$, $\State_{t+1} \sim 
\Transitions(\State'\mid  \State_t,\policy(\State_t))$. In control-design 
problems, we assume that there is a {\em cost function} $J$ on the space of
trajectories that maps each trajectory to real value. An optimal policy
$\policy^*$ is defined as the one that minimizes the expected value of the
cost function over trajectories starting from a state $\State_0$ 
sampled according to the initial distribution $\iota$.

\newcommand{\train}{{trn}}
\newcommand{\deploy}{{dpl}}

\myipara{Distribution shifts}
Obtaining an optimal control policy is often a computationally expensive 
procedure for MDPs where the underlying transition dynamics are highly 
nonlinear. Several design methods, both model-based methods such as 
model-predictive control \cite{garcia1989model}, stochastic optimal control \cite{bertsekas1996stochastic}, and model-free
methods such as data predictive control \cite{jain2017data} and deep reinforcement learning
\cite{mnih2015human}
have been proposed to solve the optimal control problem for such systems.
Regardless of whether the method is model-based or model-free, these methods
explicitly or implicitly assume a model or the distribution encoded by the 
transition dynamics of the environment. A key issue is that this distribution
may change once the system is deployed in the real-world.
To differentiate between the {\em training} environment and the 
{\em deployment} environment, we use $\Transitions_\train$ and
$\Transitions_\deploy$ to respectively denote the transition distributions.

\newcommand{\adaptedpolicy}{\hat{\policy}}
\myipara{Problem Definition} Suppose we have a system where we have
trained an optimal policy $\policy^*$ under the transition dynamics 
$\Transitions_\train$, and for a given initial state $\State_0$ sampled
from $\iota$, we sample an optimal trajectory $\tau_{opt}$ for the system using
the policy $\policy^*$. We denote this as $\tau \sim (\iota,\policy^*)$.
Let $\tau = (\State_0,\State_1,\ldots,\State_T)$. 
Let $\tau(t)$ be short-hand to denote $\State_t$. For a trajectory
that starts from the same initial state (but in the deployment environment),
we want to find the adapted policy $\adaptedpolicy$ such that the error between
$\tau_{opt}$ and the trajectory under $\Transitions_\deploy$ dynamics at time
instant $t \in [1,T]$ is small. Formally,
\begin{equation}
\label{eq:probdef}
\adaptedpolicy = \argmin_{\policy} 
    \expected_{\substack{\action_t \sim \policy(\action\mid \State_{t}), \\
    \State_{t+1} \sim \Transitions_\deploy(\State'\mid\State_t,\action_t)}} 
    \left\| \tau_{opt}(t+1) - \State_{t+1}
    \right\| 
\end{equation}

A key challenge in solving the optimization problem in \eqref{eq:probdef} is 
that $\Transitions_\deploy$ is not known. In this paper, we propose that we
learn a {\em surrogate model} for the deployment transition dynamics. Essentially,
a surrogate model is a data-driven model that approximates the actual system
dynamics reasonably accurately. There are several choices for surrogate models
including Gaussian Processes \cite{alpaydin2020introduction}, probabilistic
ensembles \cite{chua2018deep}, and deep neural networks (NN). In this paper, we focus on 
NN surrogates as they allow us to consider convex relaxations of the policy
adaptation problem.

\myipara{Surrogate-based policy adaptation} We now show that surrogate-based policy
adaptation can be phrased as a nonlinear optimization problem. First we specify the
problem of finding good surrogates. Recall that 
$\Transitions_\deploy(\State_{t+1} \mid \State_t, \action_t)$ is assumed to be a 
time-invariant Gaussian distribution with mean $\mu(\state,\action)$ and covariance 
$\Sigma(\state,\action)$. A surrogate model for the transition dynamics is a
tuple $(\mu_{NN}(\State,\action; \theta_\mu), \Sigma_{NN}(\State,\action; \theta_\Sigma))$,
where $\mu_{NN}$ and $\Sigma_{NN}$ are deep neural networks with parameters $\theta_\mu$ and 
$\theta_\Sigma$ respectively. We can train such NNs by minimizing the following 
loss functions:
\newcommand{\loss}{\mathcal{L}}
\begin{eqnarray}
    \loss_\mu(\theta_\mu) & = & \expected_{\State \sim \States, \State' \sim \Transitions_\deploy(\State'\mid \State,\action)} 
            \left\| \mu_{NN}(\State,\action; \theta_\mu) - \State' \right\| \\
    \loss_\Sigma(\theta_\Sigma) & = & \expected_{\State \sim \States} \| \Sigma_{NN}(\State, \action; \theta_\Sigma) -
            \Sigma_{s}(\State') \|
\end{eqnarray}
In the above equations, the expectation is computed by standard Monte Carlo
based sampling. In the second equation, $\Sigma_s$ represents the sample
covariance of $\State'$ w.r.t. the sample mean.

Assuming that we have learned surrogate models to a desired level of accuracy, the
next step is to frame policy adaptation as a nonlinear optimization problem. We
state the problem w.r.t. a specific optimal trajectory $\tau_{opt}$ sampled from the optimal
policy (though the problem generalizes to any optimal trajectory sampled from
an arbitrary initial state). Note that $\tau_{opt}(0) = \State_0$.
\begin{equation}\label{eq:mainopt}
 \forall t \in[0,T\!-\!1]: \action_t = \argmin_{\action \in \Actions} \| \tau_{opt}(t+1) - \mu_{NN}(\State_t,\action_t; \theta_\mu) \|
\end{equation}
We observe that as the equation above consists of a neural network, it is highly nonlinear
optimization problem. In the next section, we will show how we can convexify this 
problem.

% \noindent The basic solution strategy is to adjust the optimal action $\action_t$, generated with the autonomous agent via an optimization that aims
% to bound the difference between $\optimtrajectory$ and $\hat{\optimtrajectory}$, i.e.
% \begin{equation}\label{eq:mainopt}
%     % \resizebox{\hsize}{!}{$\maction_t=\argmin_{u} \|\State_{t+1}\ \!\!- \!\mathcal{M}_{dpl}(\dplstate_t, \! u;\paramdplfirst)\|_2,\ \State_{t+1} \!\!\sim\!  \paramtransitions_{trn}(\State_{t+1}|\State_t,\action_t).$}
%     \maction_t=\argmin_{a} \|\State_{t+1}\ \!\!- \!\mathcal{M}_{dpl}(\dplstate_t, \! a;\paramdplfirst)\|_2.
% \end{equation}
% In another word, we take $\optimtrajectory$ as a planner, and we force the deployment environment to track this planner. We approximate the distributional shift as the overall distance between $\optimtrajectory,\hat{\optimtrajectory}$ and consider the act of tracking this specific planner, as compensation for distributional shifts. The main technical contribution of our work is \textbf{to convexify optimization \eqref{eq:mainopt}} for {\em nonlinear training/deployment environments}.

\section{Policy adaptation}
\label{sec:adaptation}
% Our policy modification technique is grounded on minimization for distance between deployment environment's states $\State_t$ and the optimal trajectory, $\mathcal{T}$.
% \subsection{Solution Overview}

\myipara{Solution Overview}
The quantity in Eq.~\eqref{eq:mainopt} being minimized is at each time $t$,
the {\em residual} error between the optimal trajectory and the mean 
predicted state by the deployment environment, conditioned on its state 
and action. Let $r_{t+1} = \tau_{opt}(t+1) - \mu_{NN}(\State_t,\action_t; \theta_\mu) $.
\newcommand{\Omegainv}{\Omega}
Our main idea is:
\begin{enumerate}
    \item 
    At any given time $t$, assume that the state $\State_t$ lies in
    a confidence set described by an ellipsoid $\ellipsoid(\mu_{\State_t}, \Omegainv_{\State_t})$,
    \item
    Assume that the action $\action_t$ lies in a confidence set also described
    by an ellipsoid $\ellipsoid(\mu_{\action_t}, \Omegainv_{\action_t})$,
    \item
    Show that the residual error $r_t$ can be bounded by an ellipsoid, the center
    and shape matrix of which depends on the action $\action_t$.
    \item
    Find the action $\action_t$ that minimizes the residual error by convex optimization.
\end{enumerate}
 
We now explain each of these steps in sequence.
First, we motivate why need to consider confidence sets. Suppose
the system starts in state $\State_0$, then the state $\State_1$
is distributed according to the transition dynamics of the 
deployment environment. In reality, we are
only interested in the next states that are likely
with at least probability threshold $p$. For a multi-variate
Gaussian distribution, this corresponds to the sublevel set of
the inverse CDF of this distribution, which according to the
following lemma can be described by an ellipsoid:

\begin{lemma}\label{lem:ellips}
A random vector $r \in \mathbb{R}^n$, with Gaussian 
distribution $r \sim \gaussian(\mu, \Sigma)$, 
satisfies,
\begin{equation}\label{eq:gaussian_prob_threshold}
\Pr\left[ \frac{1}{\rho_n}(r-\mu)^\top \Sigma^{-1} (r-\mu) \leq 1 \right]=p, \quad  
\end{equation}
where,\  $
\rho_n = \Gamma^{-1}(\frac{n}{2},\frac{p}{2})$ and $\Gamma^{-1}(.,.)$ indicates the n dimensional lower incomplete Gamma function.
\end{lemma}

The above lemma allows us to define ellipsoidal confidence sets
using truncated Gaussian distributions. An ellipsoidal confidence
region with center $\mu$ and shape matrix $\rho_n\Sigma$ (where
$\rho_n$ is as defined Lemma~\ref{lem:ellips}) defines a set 
with probability measure $p$.

Now, as the policy we are considering is stochastic (which 
we also model as a Gaussian distribution), an action that 
can be taken is described by a conditional Gaussian distribution.
Let $\mu_{\action_\timeid}$ be the mean of the distribution of the
action at time $t$, then all actions with probability greater 
than $p$ can be described by a ellipsoid confidence set $\ellipsoid(\mu_{\action_t},\Omegainv_{\action_t})$. 

Because the distribution of transition dynamics may have
shifted, applying the same action $\policy^*(\State_t)$ may result 
in a residual error $r_{t+1}$ that is unacceptable. So, we want 
to find a new action, $\action_t$ which reduces the residual error. 
We assume that $\action_t$ is in an ellipsoidal uncertainty set 
by picking actions that have probability greater than a fixed
threshold $p$. We note that the center or the shape matrix of
the ellipsoidal set for the action is not known, but is a decision
variable for the optimization problem.

We note that the relation between $\action_t$ and $r_{t+1}$ is 
highly nonlinear. However, we show, how we can convexify 
this problem.

Before we present the convexification of the optimization problem,
we need to introduce the notion of the reachable set of residual
values. We call this the {\em residual reach set}. Formally,
given ellispoidal confidence region $\ellipsoid(\mu_{\State_t},
\Omegainv_{\State_t})$ for the state $\State_t$, and the ellipsoidal
confidence region $\ellipsoid(\mu_{\action_t},\Omegainv_{\action_t})$ for
the action $\action_t$, the residual reach set $\rr_{t+1}$ is defined as follows:
\begin{equation}\label{eq:resreach}
\begin{array}{ll}
\rr_{t+1}(\mu_{\action_t}, \Omegainv_{\action_t}) = \!\!
& \{ \mu_{NN}(\State_t,\action_t; \theta_{\mu}) - \tau_{opt}(t+1)
\mid \ \State_t \in \ellipsoid(\mu_{\State_t},\Omegainv_{\State_t}),\\
    &    \action_t \in \ellipsoid(\mu_{\action_t},\Omegainv_{\action_t})
\}
\end{array}
\end{equation}

In the above equation, we note that the residual 
reach set is parameterized by $\action_t$ and
$\Omega_{\action_t}$, and we wish to find the 
values for $\action_t$ and $\Omega_{\action_t}$
that minimize the size of the residual reach set.
However, the residual reach set is a non-convex
set. To make the optimization problem convex, we 
basically approximate the residual reach set by
an ellipsoidal upper bound in the set
$\ellipsoidbound(\rr_{t+1}(\mu_{\action_t},
\Omega_{\action_t}))$ (the set of all ellipsoidal
upper bounds).

We can now express the problem of finding the best adapted 
action distribution as the following optimization problem:
\begin{equation}\label{eq:bestaction}
\begin{array}{ll}
(\adaptedaction, \hat{\Omega}_{\action_t},
\hat{\Omega}_{\rr_{t+1}}) = 
& \argmin_{\mu_{\action_t},\Omega_{\action_t}} \mathbf{Logdet}(\Omega_{\rr_{\timeid+1}})\\
& \mathrm{s.t.} 
\rr_{t+1}(\mu_{\action_t},\Omega_{\action_t}) \subset  \ellipsoid(\mathbf{0}, \Omega_{\rr_{\timeid+1}}) \\ 
\end{array}
\end{equation}

Consider we set the center of ellipsoidal bound of residual reach set to be 0. This is motivated by the goal that we need to minimize the size of residuals. Equation~\eqref{eq:bestaction} selects the best
action $\maction_t \in \ellipsoid(\adaptedaction,\hat{\Omega}_{\action_t})$ s.t. the 
the ellipsoid $\ellipsoid(0,
\hat{\Omega}_{\rr_{\timeid+1}})$ is the smallest 
ellipsoid that bounds the residual reach set.

 The construction of an ellipsoidal bound over the 
 reach-set of a neural network given a single 
 ellipsoidal confidence region is derived in 
 \cite{fazlyab2019probabilistic}. The author has upgraded this
technique later for multiple ellipsoidal confidence regions in
Theorem 1 of \cite{9482752}.  We
 rephrase the key results from these papers in
 our context in Lemma~\ref{col:nn_output}.
\begin{lemma} \label{col:nn_output}
Suppose 
$\State_t \in \ellipsoid(\mu_{\State_t}, \Omega_{\State_t})$, $\action_{\timeid} \in \ellipsoid(\mu_{\action_\timeid}, \Omega_{\action_t})$. 
Then, the residual reachset $\rr_{t+1}(\mu_{\action_\timeid},{\Omega}_{\action_\timeid})$
is upper-bounded by  
$\ellipsoid(\mathbf{0}, {\Omega}_{\rr_{t+1}})$ (as defined in \eqref{eq:bestaction}) if the constraint in \eqref{eq:op_qc_assume} holds. In what follows,
$(b_\ell,W_\ell) \in \theta_\mu$ represent the bias vector and the weights of the
last layer in $\mu_{NN}$.
\begin{equation} \label{eq:op_qc_assume}
\tau_1 M_{\State_t} + \tau_2 M_{\action_\timeid} + M_{\phi} - M_{out} \leq 0, \quad \text{for some } \tau_1,\tau_2 \geq 0.
\end{equation}
Here, $M_{\phi}$ is a quadratic constraint proposed
in \cite{fazlyab2019probabilistic}, representing ReLU hidden layers in the neural network and\footnote{The parameter $N_i$ in transformation matrices $E_1, E_2$ is the number of ReLU activations in layer $i$ of $\mu_{NN}$. },
$$
\begin{aligned}
M_{\State_t}&=\frac{1}{\rho_n}E_1^\top\begin{bmatrix} -\Sigma_{\State_\timeid}^{-1} & \Sigma_{\State_\timeid}^{-1}\mu_{\State_\timeid}  \\ \mu_{\State_\timeid}^\top \Sigma_{\State_\timeid}^{-1} & -\mu_{\State_\timeid}^\top \Sigma_{\State_\timeid}^{-1}\mu_{\State_\timeid}+\rho_n  \end{bmatrix}E_1,\\
M_{\action_t}&=E_2^\top\begin{bmatrix} -\Omega_{\action_\timeid}^{-1} & \Omega_{\action_\timeid}^{-1}\mu_{\action_\timeid}  \\ \mu_{\action_\timeid}^\top \Omega_{\action_\timeid}^{-1} & -\mu_{\action_\timeid}^\top \Omega_{\action_\timeid}^{-1}\mu_{\action_\timeid}+1  \end{bmatrix}E_2
\end{aligned}
$$
$$
\begin{aligned}
E_1&=\left[\begin{matrix}  
    \begin{array}{c|c|c}  
     I_{n} \ \   0_{n \times m } &  0_{n \times \left(\sum_{i=2}^{\ell+1}N_i\right)}  & 0_{n \times 1}\\
    \hline
     0_{1 \times n+m} &  0_{1 \times \left(\sum_{i=2}^{\ell+1}N_i\right)}  & 1 \end{array}  
     \end{matrix}\right],\\
E_2&=\left[\begin{matrix}  
    \begin{array}{c|c|c}  
     0_{m \times n }  \ \ I_{m}   &  0_{m \times \left(\sum_{i=2}^{\ell+1}N_i\right)}  & 0_{m \times 1}\\
    \hline
     0_{1 \times n+m} &  0_{1 \times \left(\sum_{i=2}^{\ell+1}N_i\right)}  & 1 \end{array}  
     \end{matrix}\right]
\end{aligned}
$$
$$
M_{out} = \begin{bmatrix}C & b\\ 0 & 1\end{bmatrix}^{\top} \begin{bmatrix}-\Omega_{\rr_{\timeid+1}}^{-1} & 0 \\ 0 & 1\end{bmatrix} \begin{bmatrix}C & b\\ 0 & 1\end{bmatrix},
$$
and $C=\begin{bmatrix}0 & 0& & \cdots & W_{\ell}\end{bmatrix},\ b = b_\ell-\tau_{opt}(t+1)$, 
\end{lemma}

In the above lemma, as the adapted action and the shape matrix 
representing its covariance is assumed to be known
$M_{\action_\timeid}$ is
a fixed matrix; however, in the optimization
problem that we wish to solve, in the corresponding matrix $M_{\action_t}$, $\mu_{\action_t}$ and
$\Omega_{\action_t}$ will appear as variables, which causes
the problem to become nonlinear. We can address this by 
performing two transformations. The first transformation,
through a change of variables concentrates the nonlinearity in
a single scalar entry of $M_{\action_t}$.
 % The introduction of $\mactions_t$ as a constant ellipsoid to \cite{fazlyab2019probabilistic} and corollary \ref{col:nn_output} results in a convex constraint. However, in this study, it is proposed as a decision variable and causes nonlinearity. To linearize this constraint, we apply change of variables on $\tau_2 M_{\maction_\timeid} $ to introduce it in an applicable form to a convex optimization.
 We set $U_{\action_t}=\tau_{2} \Omega^{-1}_{\action_t},\ V_{\action_t}=\tau_{2} \Omega^{-1}_{\action_t} \mu_{\action_t}$, and the resulting $M_{\action_t}$ is shown as below:
\begin{equation}
    \begin{aligned}
    M_{\action_\timeid}^*&=E_2^\top \begin{bmatrix}
    -U_{\action_t} & V_{\action_t}\\ V_{\action_t}^\top & -\left(\tau_{2}\mu_{\action_t}^\top \Omega_{\action_t}^{-1}\right)\left(\frac{\Omega_{\action_t}}{\tau_{2}}\right)\left(\tau_{2}\Omega_{\action_t}^{-1}\mu_{\action_t}\right)+\tau_{2} 
    \end{bmatrix}E_{2}\\
    &=E_2^\top \begin{bmatrix}
    -U_{\action_t} & V_{\action_t}\\ V_{\action_t}^\top & -V_{\action_t}^\top U_{\action_t}^{-1} V_{\action_t}+\tau_{2} 
    \end{bmatrix}E_2
    \end{aligned}
\end{equation}
The proposed matrix, $M_{\action_\timeid}^*$, is nonlinear where the nonlinearity shows up in the scalar variable $V_{\action}^\top U_{\action}^{-1} V_{\action}$. 
% In the matrix $M_{\maction}$, nonlinearity shows up through the term $V_{\maction}^\top U_{\maction}^{-1} V_{\maction}$ which is a scalar and concentration of nonlinearities in a scalar value is beneficial in convexification of the solution as is described later. 

We also note that the adapted actions should satisfy actuator
bounds $[\mathbf{\ell},\mathbf{u}]$, we include this as a convex constraint below: 
% The modified policy, $\maction_\timeid$ should respect the bounds of autonomous agent $\action_\timeid \in [ \vec{\ell}, \vec{u} ],\ \vec{\ell}, \vec{u} \in  \mathbb{R}^m$. Thus, we also add a new constraint on modified policy regarding the actuator bounds. We propose this bound in lemma \ref{lem:actuatorbound}, and we defer the proof to the appendix,

% \begin{lemma} \label{lem:actuatorbound}
% The actuator bound's constraint can be proposed in a convex form as,
\begin{equation}\label{eq:actuatorboundconstraint}
    U_{\action_t}\mathbf{\ell} \leq V_{\action_t}\leq U_{\action_t}\mathbf{u}, 
\end{equation}
and we defer the proof to the appendix B. 
Before stating the final theorem, we make an observation
about \eqref{eq:bestaction}. Without additional constraints,
the optimal solution to \eqref{eq:bestaction} always returns
${\hat{\Omega}}_{\action_\timeid}$ such that
$\mathsf{tr}(\hat{\Omega}_{\action_\timeid})=0$ (proof
in the appendix A). This causes
numerical errors, as the inverse of this matrix is appeared in Lemma \ref{col:nn_output} which is not defined.
To avoid such a problem, we impose a tiny lower bound on 
the trace of this matrix.

% This  implies $\maction_\timeid$ is a deterministic variable and 
% $\mactions_\timeid=\left\{\maction_\timeid\right\}$ is a singleton.

% The proof is included in Appendix A. As we show later in Corollary \ref{col:nn_output} the term $\Omega_\timeid^{-1}$ appear in the optimization problem. Therefore, Lemma \ref{clm:singleton} indicates a problematic computational complexity in optimization. We describe it further in detail in the proof of Theorem \ref{thm:action_opt}. To address this issue, we introduce a new constraint by provision of an infinitesimal lower bound for size of $\mactions_t$ to avoid a singleton.

 Finally, given that all the constraints for the optimization of $\ellipsoid(\mu_{\action_t}, \Omega_{\action_t})$ are provided, we can collect them in a convex optimization that results in  the modified action set. The following Theorem characterizes the correctness of the modified action and its conservatism. We defer the proof to the appendix.
\begin{theorem}\label{thm:action_opt}
 Given the regulation factor $\delta>0$, and defining $\Omega = \Omega_{\mathcal{R}_{\timeid+1}}^{-1}$, assume decision variables $\tau_1,\tau_2 \geq 0$. Then the following convex optimization,
\begin{equation} \label{eq:action_optm}
\left\{\begin{aligned}
&\min_{M_\phi, V_{\action_t}, U_{\action_t}, \tau_1, \tau_2}  -\mathbf{Logdet}(\Omega)\\
% \vspace{-2mm}
\ \ &\mathbf{s.t.}\  -\tau_1 M_{\State_\timeid}-E_2^\top \begin{bmatrix}
    -U_{\action_t} & V_{\action_t}\\ V_{\action_t}^\top & \tau_{2}\end{bmatrix}E_2- M_{\phi} + M_{out}\geq 0,\\ 
    & U_{\action_t}\vec{\ell} \leq V_{\action_t}\leq U_{\action_t}\vec{u}  , \quad \mathbf{tr}(U_{\action_t}) \delta \leq \tau_2.
\end{aligned}\right.
\end{equation}
results in  values $(V_{\action_t},\ U_{\action_t})$ such that the modified deterministic decision $\maction_t^c$ can be approximated with $\maction_t^c=\adaptedaction=U_{\action_t}^{-1}V_{\action_t}$. 
% with a small confidence region $\mactions_t=\mathcal{B}(U_{\maction}^{-1}V_{\maction},\ \tau_{2} U_{\maction}^{-1} )$. 
\end{theorem}
% \begin{proof}
 % see appendix. The contribution of distributional shift, $\delta_d$, in this result is currently implicit. We analyse it in detail in future works and formalize it in mathematical form. 
% \end{proof}
Regarding the possible robustness issues with models (adversarial examples) we need to make a comparison between $\pi^*$ and ${\maction}_t^c$ as a precautionary measure and select for the best choice for the modified action $\maction_t$, via,
\begin{equation}\label{eq:comparison}
\maction_t = \underset{a \in \left\{\pi^*,\ {\maction}_t^c\right\}}{\arg\min} \| \tau_{opt}(t+1)-\mu_{NN}(\State_t,a;\theta_\mu)\|_2.
\end{equation}
See Appendix D for more detail. We summarize the main steps of our proposed method in  algorithm \ref{alg_main_algorithm}. 
\subsection{Scalability} \label{sec:scalable}
The conservatism of tight ellipsoidal bound approximation introduced in \cite{fazlyab2019probabilistic} increases with the complexity of neural network's structure and results in inaccurate solution for Theorem \ref{thm:action_opt}. However, for a highly nonlinear deployment environment, this is necessary to train a deep neural network for the surrogate. In response to this problem, (similar to \cite{lusch2018deep}), we utilize an embedder network, $\mathcal{M}_p$, which maps the state $\State_t$ to another space $\State_\timeid' \in \mathbb{R}^{n'}$ $\left(\State_t'=\mathcal{M}_p(\State_t; \param_p)\right)$, such that $\State_\timeid'$ is more tractable than $\State_\timeid$ for training purposes. We next define the surrogate model and it's parameters $\theta_\mu$ based on $\State_t'$ as,
$$
\mu_{\State_{\timeid+1}} = \mu_{NN}(\State_\timeid',\ \maction_{\timeid}; \theta_\mu).
$$
Given this setting for a highly nonlinear deployment environment, the neural network $\mu_{NN}$ is not necessarily a deep neural network. Thus, given the pair $(\State_\timeid, \action_\timeid)$ as the input vector and $\State_{\timeid+1}$ as output, we  arrange a training procedure for the function,
$$
\mu_{\State_{\timeid+1}} = \mu_{NN}\left(\mathcal{M}_p(\State_t; \param_p),\ \action_{\timeid}; \theta_\mu \right)
$$
to learn the parameters $\theta_\mu, \param_p$ together and utilize $\theta_\mu$ in the convex programming. In another word, given the distribution of $\State_\timeid$ and the parameters $\param_p$, we can approximate the distribution for $\State'_\timeid$ with Gaussian mixture  model techniques \cite{reynolds2009gaussian} to introduce its confidence region to the convex optimization (through $\mu_{NN}$) for  policy modification.

% such that $\State_\timeid'$ is \textbf{trainable with a shallow neural network}. Fig.\ref{fig:EmbedderDNN} shows the proposed two-step structure for training the shallow surrogate model for the deployment environment in highly nonlinear environments. 
\begin{algorithm*}[t]
\SetKwInput{KwData}{Input}
\SetAlgoLined
\DontPrintSemicolon
\KwData{$\State_1$ and trained autonomous agent, $\pi^*$}
\KwResult{Small reachset for residual with its center closer to origin.}
{ $\bullet$ Sample the optimal trajectory $\tau_{opt}$.}\;
\ForEach{time step $t$}{
{  $
\bullet \left\{
\begin{aligned} 
&\textbf{if} (t=1) \quad \left[\maction_1 \gets \pi^*\right] \\
&\textbf{else}  (t\geq 2) \left\{
\begin{aligned} 
& \text{1- Apply Theorem \ref{thm:action_opt} and compute } \maction_t^c \\
& \text{2- Select the best action between } \maction_t^c \text{ and } \pi^* \text{, with \eqref{eq:comparison} and return } \maction_t.
\end{aligned}\right.
\end{aligned}\right.$}\;
{$\bullet$ Employ the observation $\State_t$ and $\maction_t$ to characterize the confidence region $\States_{t+1}$ Using the deep surrogate (see Appendix D) for the deployment environment.}\;
{$\bullet$ Record the observation $\State_{t+1}$ generated by exertion of $\maction_t$ to the environment}\;
}
\caption{ Compensation Process for Distributional shifts}
\label{alg_main_algorithm}
\end{algorithm*}

\section{Experimental Results}
\label{sec:results}
\myipara{Comparison with PSO} We assume simple car environment and compare the performance of our convex programming technique with Particle Swarm Optimization,\ (PSO) \cite{kennedy1995particle}\footnote{While it is well-known that nonlinear optimization techniques lack guarantees and can suffer from local minima, techniques like particle swarm work well in practice, especially in low-dimensional systems. Hence, we perform this comparison to show that convexification outperforms state-of-the-art global optimization approaches to residual minimization.}, on solving the optimization \eqref{eq:mainopt}. PSO has shown acceptable performance in low dimensional environments. Thus, we plan to show our convex programming technique can outperform PSO even if the scalability is not an important issue. 
% In order to show the accuracy of the proposed technique we compare our result with particle swarm optimization algorithm (PSO) \cite{kennedy1995particle}\footnote{While it is well-known that nonlinear optimization techniques lack guarantees and can suffer from local minima, techniques like particle swarm work well in practice especially in low-dimensional systems. Hence, we perform this comparison to show that convexification outperforms state-of-the-art global optimization approaches to residual minimization.} which has an acceptable performance on low dimensional systems. We address the simple car problem with three states and one control input and we propose two alternate methods for optimization: a baseline method that solves our original nonlinear optimization \eqref{eq:mainopt} directly utilizing particle swarm optimization algorithm and the proposed policy adaptation method that utilizes a convex programming to solve for \eqref{eq:mainopt}.
The environment represents the following simple car model:
\begin{equation}
\begin{aligned}
    &\dot{x}= u \mathbf{cos}(\theta), \ \ 
    \dot{y}= u \mathbf{sin}(\theta), \ \
    \dot{\mathbf{sin}}(\theta)= \frac{u}{\ell}\mathbf{tan}(\phi) \mathbf{cos}(\theta), \\
    &\dot{\mathbf{cos}}(\theta)= -  \frac{u}{\ell}\mathbf{tan}(\phi) \mathbf{sin}(\theta)
\end{aligned}
\end{equation}
The system represents a car of length $\ell$ moving with constant velocity $u$ and driven with control action $\phi$. The training environment is characterized by $\ell=2.5$ and, $u=4.9$ while the deployment environment is slightly different with $\ell=2.1$ and $u=5.1$. We collect a training data set from deployment environment with time step $0.01$ second.

\begin{figure}
\hspace{-20mm}    \includegraphics[width=1.3\textwidth]{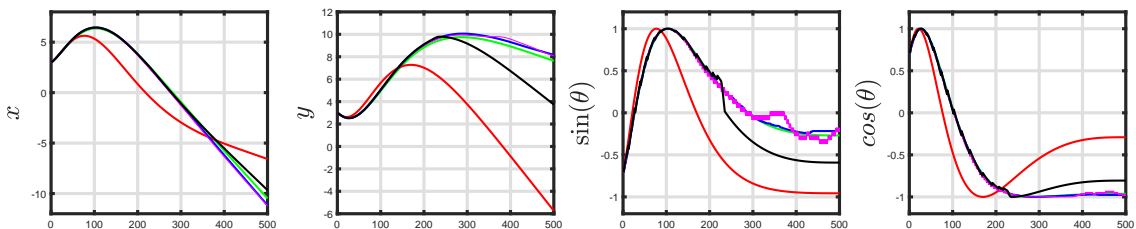}
    \caption{Shows the comparison between PSO and our convex programming. The green and blue curves are the results of algorithm \ref{alg_main_algorithm} and optimal trajectory, respectively 
    . The red curves represent the deployment environment's trajectory when there is no policy modification. We utilize PSO for optimization \eqref{eq:mainopt} on the same model with convex programming, $\mu_{NN}$ where the resultant trajectory is demonstrated in black. We also utilize PSO over the deep model, $\mu_{NN}^*$ and the resultant trajectory is demonstrated in  magenta.}
    % \vspace{-3mm}
    \label{fig:comparison}
\end{figure}

% Since the model may suffer from robustness issues such as adversarial examples, we must compare performance of theorem \ref{thm:action_opt}, ($\maction_\timeid$) with the autonomous agent's decision $\action_\timeid$. 
We train two surrogates for the deployment environment $\mu_{NN}, \mu_{NN}^*$ from deployment environment. The former is utilized in optimization \eqref{eq:action_optm} and is a ReLU neural network with dimension $[5,8,4]$. This ReLU neural network is obtained from the proposed procedure in section \ref{sec:scalable}. The latter is utilized for comparison discussed in equation \eqref{eq:comparison} and Appendix D, which is a deep $\mathbf{tanh}()$ neural network of dimension $\left[5, 200, 200, 200, 200, 200, 200, 200, 4 \right]$.
% Since particle swarm algorithm does not support model uncertainty we assume a small uncertainty to have a valid comparison between our convex programming and PSO. 
The results of policy modification are presented in Fig.\ref{fig:comparison}. This figure presents the optimal trajectory, $\tau_{opt}$, in green color. This trajectory is simulated with an optimal control and hight-fidelity surrogate for the training environment computed from a model-based algorithm. The blue curves are the results of algorithm \ref{alg_main_algorithm} for (500 steps), which closely tracks the optimal trajectory in all the three states $x,y,\theta$. The red curves represent the deployment environment's trajectory when there is no policy modification. The run time for convex programming is between$[0.005,0.027]$ on a personal laptop with YALMIP and MOSEK solver. Thus, we restrict the run time of PSO with $0.027$ and employ it for policy modification. In one attempt, we utilize PSO for optimization \eqref{eq:mainopt} on the same model with convex programming, $\mu_{NN}$ where the resultant trajectory is demonstrated in black. In another attempt we utilize PSO over the deep model $\mu_{NN}^*$ and the resultant trajectory is demonstrated in  magenta. The results clearly shows our convex programming outperforms the PSO in both cases.

\begin{figure}
\hspace{-20mm}    \includegraphics[width=1.3\textwidth]{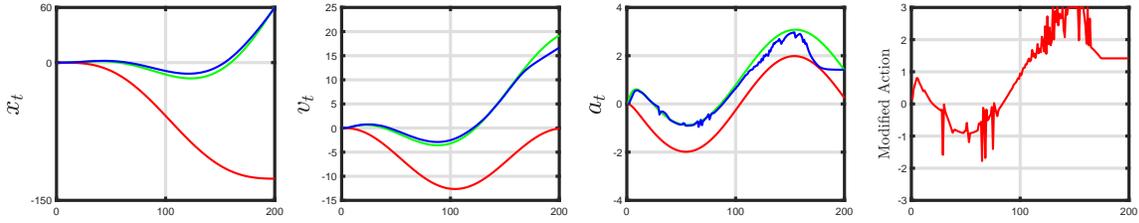}
    \caption{Shows the results of policy modification on stochastic linear environment of a car. In this figure, the green curve presents the simulated optimal trajectory.  Blue and red curves also represent the trajectory of deployment environment in the presence and absence of policy modification, respectively.}
    % \vspace{-3mm}
    \label{fig:linexample}
\end{figure}

\myipara{Linear Environment of a Car} The training environment is a stochastic linear dynamics as follows:
$$
\begin{aligned}
\mathbf{x}_{t+1}&=\begin{bmatrix} 1 & 0.1 &0.0047\\0 & 1 & 0.0906 \\ 0& 0& 0.8187\end{bmatrix} \mathbf{x}_t + \begin{bmatrix} 0.003 \\ 0.0094\\ 0.1813 \end{bmatrix} u_t + \nu_t,\\
\nu_t &\sim \gaussian\left(\begin{bmatrix}0 & 0 & 0.2 \end{bmatrix}^\top, \exp(-8)I_3 \right)
\end{aligned}
$$
The deployment environment is also a stochastic linear dynamics as follows:
$$
\begin{aligned}
\mathbf{x}_{t+1}&=\begin{bmatrix} 1 & 0.1 &0.0046\\0 & 1 & 0.0885 \\ 0& 0& 0.7788\end{bmatrix} \mathbf{x}_t + \begin{bmatrix} 0.004 \\ 0.0115\\ 0.2212 \end{bmatrix} u_t + \eta_t,\\
\eta_t &\sim \gaussian\left(\vec{0}, \exp(-8)I_3 \right)
\end{aligned}
$$
where the sampling time is $t_s=0.1\ s$. The state $\mathbf{x}_t \in \mathbb{R}^3$ is defined as $\mathbf{x}_t=\left[ x_t,\  v_t,\  a_t \right]^\top$ that are position, velocity and acceleration of car respectively. The scalar action $u_t$ is also bounded within $u_t \in[-3,\ 3]$. Since the environment is linear, it is not required to use the embedder network. Thus, we train only one  surrogate for the deployment environment $\mu_{NN}$ with a 2 hidden layer ReLU neural network of dimension $[4,10,5,3]$. We also have access to the model of training environment and a trained optimal feedback policy. Therefore, we perform policy modification through algorithm \ref{alg_main_algorithm} for deployment environment and the results are presented in Fig.\ref{fig:linexample}. In this figure, the green curve presents the simulated optimal trajectory.  Blue and red curves also represent the trajectory of deployment environment in the presence and absence of policy modification, respectively. This figure shows the algorithm \ref{alg_main_algorithm} forces the deployment environment to track the planner $\tau_{opt}$ and the policy modification process is successful. 

\begin{figure}[t]
   \hspace{-15mm} \includegraphics[width=1.2\textwidth]{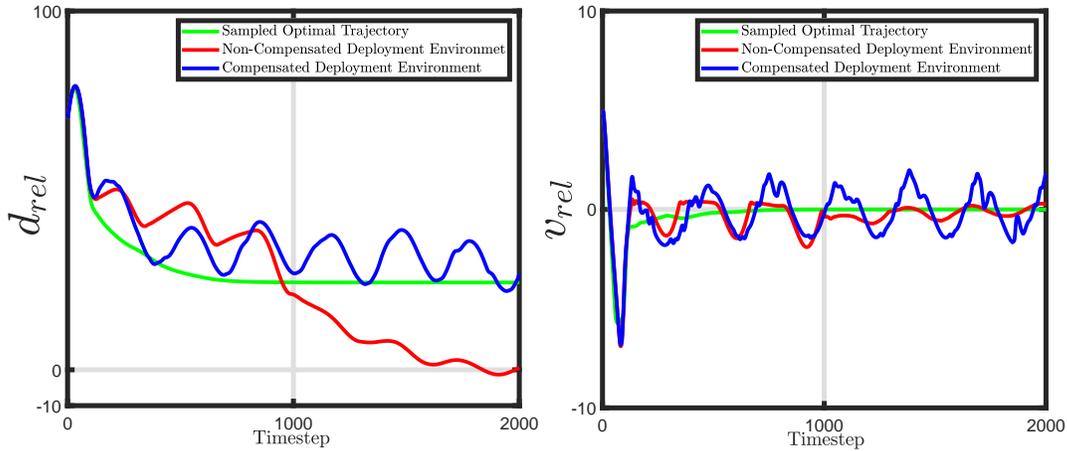}
    \caption{The green curves represent the optimal trajectory for $v_{rel}$ and $d_{rel}$, while the red and blue curves present the trajectory of deployment environment without policy adaptation and with adaptation, respectively.}\label{fig:cruiseresult}
\end{figure}
\myipara{Adaptive Cruise Control} Consider the Simulink environment for adaptive cruise control in MATLAB documentation \footnote{https://www.mathworks.com/help/reinforcement-learning/ug/train-ddpg-agent-for-adaptive-cruise-control.html}. We consider this trained feedback controller and assume we have access to the model of training environment. We then simulate the optimal trajectory $\tau_{opt}$ with model and controller. This controller is trained over $14$ hours, which clearly shows how learning a new controller can be expensive and justifies the contribution of our technique. The input of the trained controller is the vector $\mathbf{x}=\small{[\int v_{err},\ v_{err}, \ v_{ego} ]^\top}$. Thus, we take this vector as the state of the environment \footnote{Here $v_{err}$ is a logic based function of $x_{ego},x_{lead},v_{ego}$ and $v_{lead}$. See the MATLAB documentation for more detail. Here, $(v_{ego},v_{lead})$ and $(x_{ego},x_{lead})$ are the velocity and position for ego and lead car, respectively.}. 
% We next consider an adaptive cruise control environment with a pre-trained agent. We utilize the high fidelity-model to sample the optimal trajectory $\optimtrajectory$ (planner). We consider the input signal to the agent as the state of the environment,
% % \begin{equation}\label{eq:cruisestate}
% $\small{\State_t=[\int v_{err},\ v_{err}, \ v_{ego} ]^\top}$.
% % \end{equation}
This implies lead-car, ego-car and signal processing block are all together the environment. Consider, this environment is highly nonlinear due to the presence of logic based relations in the signal processing block. The implemented value of $V_{set}$ on the training environment is set on $30\ m/s$ while it is mistakenly set on $34.5\ m/s$ in the deployment environment. This difference characterizes the distributional shift.

% The next parameter that contributes to the distributional shift is the dynamics of ego and lead car. The parameter $\sigma$ for ego and lead cars in training environment is $\sigma=0.5$. On the other hand, this parameter in the deployment environment is $\sigma=0.45$ for lead and $\sigma=0.55$ for ego car.
% Consider parameters $x_{ego}$ and $v_{ego}$ are function of state, and their optimal trajectory in Fig.\ref{fig:cruiseresult} are computed based on $\mathcal{T}$. Therefore, if we compensate for distributional shift and keep $\dplstate_t$ close to $\State_t$, then $d_{rel}=x_{lead}-x_{ego}$ and $v_{rel}=v_{lead}-v_{ego}$ in deployment and training environment must remain close. 
Fig.\ref{fig:cruiseresult} shows the evolution of relative velocity and relative position, $v_{rel}, d_{rel}$ between lead and ego cars. The green line shows the simulation for $v_{rel}, d_{rel}$ when the optimal policy, $\pi^*$ is applied on the training environment. On the other hand, blue and red lines show the evolution of $v_{rel}, d_{rel}$ in the presence and absence of policy modification, respectively. Policy modification process aims to force the states of deployment environment to track optimal trajectory $\tau_{opt}$. Consider the parameter $d_{rel}<0$ on red line at time $t=183 \mathbf{s}$. Thus, the distributional shift leads to accident in the absence of policy modification. Fig.\ref{fig:cruiseresult} shows, our policy modification technique secures the system against the possible accident. 
% In this problem, We considered $\Sigma_{\dplstate_{t+1}|\dplstate_t,\maction_{t}} = 10^{-8}I$.

% \begin{figure}
%     \includegraphics[width=\linewidth]{figures/Cruiseresulthoriz.eps}
%     \caption{\small{The green curves represent the optimal trajectory for $v_{rel}$ and $d_{rel}$, while the red and blue curves present the trajectory of deployment environment without policy adaptation and with adaptation respectively.}} \label{fig:cruiseresult}
% \end{figure}

\vspace{-1mm}
\section{Conclusion and Related Work }
\vspace{-1mm}
\label{sec:conclusion}
\vspace{-1mm}
\myipara{Related work} In \cite{clavera2017policy}, the authors separate  the  learned policy from  the  raw  inputs  and  outputs to  ease  the  transfer from  simulation. In \cite{christiano2016transfer} a method to use a learned deep inverse dynamics model to decide which real-world action is most suitable to achieve the same state as the simulator is proposed. Mutual alignment transfer learning approaches employ auxiliary rewards  for transfer learning under discrepancies in system dynamics for simulation to robot transfer (\cite{wulfmeier2017mutual}).  Approaches such as \cite{eysenbach2020off}
compensate for the difference in dynamics by modifying the reward function such that the modified reward function penalizes the agent for visiting states and taking actions in the source domain which are not possible in the target domain.  In \cite{pinto2017robust}, the authors study robust adversarial reinforcement learning. Inspired with the $\|H\|_\infty $ control idea, they assume the destabilizing adversaries like the gap between simulation and environment as uncertainties and devise a learning algorithm which considers the worst case adversary and is robust against it. Transfer learning has also been investigated in the multi-agent setting \cite{li2019robust}, where the problem of training agents with continuous actions is studied to ensure that the trained agents can still generalize when their opponent's policies alter. \cite{paul2019fingerprint} proposed to  use  Bayesian  optimization (BO) to actively select the distribution of the environment variable that maximizes the improvement generated by each iteration of the policy gradient method. Unlike the authors of \cite{eysenbach2020off} who propose a reward modification technique, in this work we propose a policy modification technique to tackle the problem when the environment model in training is different from what is expected.
\vspace{-1mm}

\myipara{Conclusion \& Broader Impact} In this work, we presented a linearization algorithm for a non-linear system trained by deep neural networks equipped with ReLU activation
function and proposed a convex optimization based framework to do the
distributional shift compensation on an unknown model with unknown
distributional shift. The benefit is the convenience of computation and ability of the controller to respond to the distributional shift instantaneously, with a small cost of added conservatism due to the ellipsoidal bound based linearization. 
\vspace{-1mm}

\myipara{Future work} Algorithm \ref{alg_main_algorithm} is currently running the optimization step by step that is a greedy method. We plan to derive a single optimization that covers all the trajectory to improve the result.

\bibliographystyle{splncs04}
\bibliography{main}

\section{Appendix}
\label{sec:appendices}
\section*{Appendix A: Generic computational Error}\label{apdx:whyzero}
We present a brief summary of \cite{fazlyab2019probabilistic} and \cite{9482752} in Appendix E and here we directly focus on the necessary steps for the proof. We denote $\mathcal{M}_t : (\States \times \actions) \to \rr_{\timeid+1}$ as the neural network where represents the residual. This network is equivalent with $\mu_{NN}$ but the last bias vector is shifted with $\tau_{opt}(t+1)$. Assume $\mathcal{R}_{\timeid+1}, \mathcal{R}_{\timeid+1}^c$ are the residual reachsets when $(\States_\timeid \times \actions_t)$ and $(\States_\timeid \times \actions_t^c)$ are introduced in, $\mathcal{M}_\timeid$ respectively ($\actions_t^c \subset \actions_t$). This implies, if $\action_t \in \actions_t^c$, then $\action_t \in \actions_t$ and therefore, for $\actions_\timeid = \mathcal{E}(\mu_{\action_\timeid}, \Omega_{\action_t}),$ 
\vspace{-1mm}
$$
\begin{bmatrix}\action_t\\1\end{bmatrix}^\top \begin{bmatrix} -\Omega_{\action_t}^{-1} & \Omega_{\action_t}^{-1}\mu_{\action_\timeid}  \\ \mu_{\action_\timeid}^\top \Omega_{\action_t}^{-1} & -\mu_{\action_\timeid}^\top \Omega_{\action_t}^{-1}\mu_{\action_\timeid}+1  \end{bmatrix} \begin{bmatrix}\action_t\\1\end{bmatrix} \geq 0, 
$$
\vspace{-1mm}
which suffices to say, regarding the optimization \eqref{eq:fazlyab}, the optimal upper-bound for the $\rr_{\timeid+1}$ produced by $(\States_\timeid \times \actions_t)$ is in fact a feasible solution for the upper-bound of 
$\mathcal{R}_{\timeid+1}^c$ obtained from $(\States_\timeid \times \actions_t^c)$ (see the summary of \cite{fazlyab2019probabilistic,9482752} in Appendix E). This implies the optimal objective function of optimization \eqref{eq:fazlyab} in the second problem (taking $(\States_\timeid \times \actions_t^c)$ as input) is less than the optimal objective function of optimization\eqref{eq:fazlyab} in the first problem (taking $(\States_\timeid \times \actions_t)$ as input). Therefore, we conclude  $\mathbf{Logdet}(\Omega_{\mathcal{R}_{t+1}^c}) < \mathbf{Logdet}(\Omega_{\mathcal{R}_{t+1}})$. In another word, we conclude, replacing $\actions_\timeid$ with $\actions_\timeid^c$ results in smaller upper-bound on the residual reachset.

Now assume the optimal confidence region for modified action, $\actions_\timeid$ in optimization \eqref{eq:bestaction} contains nonempty subset, $\actions_\timeid^c \subset \actions_\timeid$. However, we know the ellipsoidal bound of residual reachset is still reducible  by replacing $\actions_\timeid$ with $\actions_\timeid^c$. This is a contradiction because we have already concluded $\actions_\timeid$ results in smallest upper-bound for the residual reachset. Thus, the optimal confidence region $\actions_\timeid$ contains no subset and is a singleton. In another word $\mathrm{tr}(\Omega_{\action_t})=0$.

\section*{Appendix B: Construction of Actuator Bound}\label{apdx:actionbound}

The proposed action should be inside the following hyper-rectangle:
$\vec{\ell} \leq \action_t \leq \vec{u}$. In Appendix A we proved that the solution of optimization \eqref{eq:action_optm} is a singleton  $\actions_t=\left\{\action_t\right\}$, therefore we neglect the shape matrix, $\tau_2 U_{\action_t}^{-1}$ and only bound the mean value in the mentioned hyper-rectangle, $\vec{\ell} \leq U_{\action_t}^{-1}V_{\action_t} \leq \vec{u}$, which implies,
$U_{\action_t}\vec{\ell} \leq V_{\action_t} \leq U_{\action_t}\vec{u}$
\section*{Appendix C. Proof of Theorem \ref{thm:action_opt}:} 
Based on \cite{fazlyab2019probabilistic} we know the sufficient condition for an ellipsoid $\mathcal{E}(0,\Omega)$ to bound the reachset of the residual is,
\begin{equation}\label{eq:1}
\tau_1 M_{\State_\timeid}+E_2^{\top} \begin{bmatrix}
    -U_{\action_t} & V_{\action_t}\\ V_{\action_t}^\top & -V_{\action_t}^\top U_{\action_t}^{-1}V_{\action_t}+\tau_{2}\end{bmatrix}E_2+ M_\phi - M_{out}\leq 0
\end{equation}
we move the linear terms to the right and keep the nonlinear term at the left,
\begin{equation}\label{eq:2}
\begin{aligned}
E_2^{\top} \begin{bmatrix}
    0 & 0\\ 0 & -V_{\action_t}^\top U_{\action_t}^{-1}V_{\action_t}\end{bmatrix}E_2 \leq& -\tau_1 M_{\State_\timeid}-E_2^{\top} \begin{bmatrix}
    -U_{\action_t} & V_{\action_t}\\ V_{\action_t}^\top & \tau_{2}\end{bmatrix}E_2\\
    &- M_\phi + M_{out}
    \end{aligned}
\end{equation}\label{eq:3}
The matrix in the left of inequality is negative definite, therefore if we introduce the new constraint,
\vspace{-1mm}
\begin{equation}\label{eq:4}
-\tau_1 M_{\State_\timeid}-E_2^{\top} \begin{bmatrix}
    -U_{\action_t} & V_{\action_t}\\ V_{\action_t}^\top & \tau_{2}\end{bmatrix}E_2- M_\phi + M_{out} \geq 0
\end{equation}\vspace{-1mm}
we have satisfied the required constraint in \eqref{eq:2}. Based on our observations, this new linear constraint will not impose conservatism because the value $V_{\action_t}^\top U_{\action_t}^{-1}V_{\action_t}$ is always near to zero, thus we are neglecting the negative value of the nonlinear term.  As we discussed before, we know $\Omega_{\action_t}=\tau_2 U_{\action_t}^{-1}$ converges to zero, this implies there is a chance for $U_{\action_t}$ to become unbounded and this is why an infinitesimal $\Omega_{\action_t}$ is problematic for our convex optimization. In order to avoid unbounded solution for $U_{\action_t}$, we provide a small lower bound on the $\mathrm{tr}(\Omega_{\action_t})$. We know $\Omega_{\action_t}$ is a positive definite matrix, therefore if $\mathrm{tr}(\Omega_{\action_t}^{-1})$ is smaller than a big number, $\sigma$, that suffices to have  $\mathrm{tr}(\Omega_{\action_t})$ to be greater than a small number (lower bound on size of $\actions_t$). This can be rephrased with the convex constraint $\mathrm{tr}(U_{\action_t})\leq \tau_2 \sigma$, or in another word, $ \mathrm{tr}(U_{\action_t})\delta \leq \tau_2 $ where $\delta= \frac{1}{\sigma}$ is preferably a small number. Thus, to justify the presence of convex constraint $\mathrm{tr}(U_{\action_t}) \delta \leq \tau_2$, we mention this is just a precautionary measure ($\delta$ is very small) to avoid unbounded solutions. 

\section*{Appendix D. Comparison} \label{apdx:comparison}

Due to presence of adversarial examples, we need to certify the modified action performs better than autonomous agent on deployment environment. Since we can not utilize the environment directly for this purpose, we must employ a deep surrogate model for deployment environment. On the other hand, reading through \cite{fazlyab2020safety} clarifies, although a deep network is accurate, it results in noticeable conservatism for convex programming in Theorem \ref{thm:action_opt}. Therefore, we train two networks for surrogate in a highly nonlinear environment. The former will be obtained from Embedded technique in section \ref{sec:scalable} and will be utilized for convex programming. The latter is a very deep network that provides reliability for an accurate comparison between, $\pi^*$ and $\action_\timeid^c$. We call this deep neural network as, $\mu_{NN}^*$.    

\section*{Appendix E. Brief summary of \cite{fazlyab2019probabilistic}}

We present the solution summary with parameters of our specific problem for more clarification. Assume the confidence regions for state and actions $\State_\timeid \in \States_\timeid,\ \action_\timeid \in \actions_\timeid$ are fixed and known. Then the tool provided in \cite{fazlyab2019probabilistic,9482752} proposes a convex optimization for tightest ellipsoidal upper-bound over residual's reachset $\mathcal{R}_{\timeid+1}$. In this research, we add another constraint and fix the center of the mentioned upper-bound on the origin to make it certain that the residual decreases in Euclidean norm. Therefore, the tool \cite{fazlyab2019probabilistic,9482752} is utilized to present the tightest upper-bound such that, $\mathcal{R}_{\timeid+1} \subset \mathcal{E}(\vec{0}, \Omega_{\mathcal{R}_{\timeid+1}}^{-1})$. We know $\action_\timeid \in \actions_\timeid:= \mathcal{E}(\mu_{\action_t}, \Omega_{\action_t})$ therefore:
\vspace{-1mm}
\begin{equation}\label{eq:1_1}
\begin{bmatrix}\action_t\\1\end{bmatrix}^\top \underline{\begin{bmatrix} -\Omega_{\action_t}^{-1} & \Omega_{\action_t}^{-1}\mu_{\action_\timeid}  \\ \mu_{\action_\timeid}^\top \Omega_{\action_t}^{-1} & -\mu_{\action_\timeid}^\top \Omega_{\action_t}^{-1}\mu_{\action_\timeid}+1  \end{bmatrix} }_{Q_1}\begin{bmatrix}\action_t\\1\end{bmatrix} \geq 0
\end{equation}
We also know $\State_\timeid \in \States_\timeid:= \mathcal{E}(\mu_{\State_\timeid}, \rho_n \Sigma_{\State_\timeid})$ therefore:
\begin{equation}\label{eq:2_1}
\frac{1}{\rho_n} \begin{bmatrix}\State_t\\1\end{bmatrix}^\top \underline{\begin{bmatrix} -\Sigma_{\State_\timeid}^{-1} & \Sigma_{\State_\timeid}^{-1}\mu_{\State_\timeid}  \\ \mu_{\State_\timeid}^\top \Sigma_{\State_\timeid}^{-1} & -\mu_{\State_\timeid}^\top \Sigma_{\State_\timeid}^{-1}\mu_{\State_\timeid}+\rho_n  \end{bmatrix}}_{Q_2}\begin{bmatrix}\State_t\\1\end{bmatrix} \geq 0,
\end{equation}
In the next attempt \cite{fazlyab2019probabilistic} suggests us to concatenate all the post-activations, in the residual's model $\mathcal{M}_\timeid$ and generate vector $\mathbf{x}=[z^{1\top},z^{2\top}, \cdots,z^{\mathcal{L}-1\ \top}]^\top$. Then they propose a symmetric matrix $Q_\phi$ which satisfies the quadratic constraint,
\vspace{-2mm}
\begin{equation}\label{eq:3_1}
\begin{bmatrix}\mathbf{x}\\1\end{bmatrix}^\top Q_\phi \begin{bmatrix} \mathbf{x}\\1\end{bmatrix} \geq 0.
\end{equation}
The ultimate goal is to prove the residual $\Delta_{\timeid+1} \in \mathcal{E}(\vec{0}, \Omega_{\mathcal{R}_{\timeid+1}})$. Therefore, defining $\Omega = \Omega_{\mathcal{R}_{\timeid+1}}^{-1}$ we should propose a constraint that implies,
\vspace{-2mm}
\begin{equation}\label{eq:4_1}
\begin{bmatrix}r_{\timeid+1}\\1\end{bmatrix}^\top \begin{bmatrix} -\Omega & 0  \\ 0 &  1  \end{bmatrix} \begin{bmatrix}r_{\timeid+1}\\1\end{bmatrix} \geq 0
\end{equation}
To propose such a constraint authors in \cite{fazlyab2019probabilistic} suggest defining the base vector $\mathbf{z}= [\State_\timeid^\top,\ \action_\timeid^\top,\ \mathbf{x}^\top, 1]^\top$ and define the linear transformation matrices $E_1, E_2, E_3$ and matrix $C$ as,
$$
\begin{bmatrix}\State_\timeid \\1\end{bmatrix}= E_1 \mathbf{z},\  \begin{bmatrix}\action_\timeid \\1\end{bmatrix}= E_2 \mathbf{z},\ 
\begin{bmatrix}\mathbf{x} \\1\end{bmatrix}= E_3 \mathbf{z}, \ \begin{bmatrix}r_{\timeid+1}\\1\end{bmatrix} = \begin{bmatrix}C & b\\ 0 & 1\end{bmatrix} \mathbf{z},
$$
and $C=\begin{bmatrix}0 & 0& & \cdots & W_{\ell}\end{bmatrix}$, $b= b_\ell -\tau_{opt}(t+1)$, $(b_\ell,W_\ell) \in \theta_\mu$ represent the bias vector and the weights of the
last layer in $\mu_{NN}$, and add the left side of equations \eqref{eq:1_1}, \eqref{eq:2_1}, \eqref{eq:3_1} which provides the following inequality:
\begin{equation}
    \mathbf{z}^\top \left( \tau_1 \underline{E_1^\top Q_1 E_1}_{M_{\State_\timeid}} + \tau_2 \underline{E_2^\top Q_2 E_2}_{M_{\action_\timeid}} + \underline{E_3^\top Q_\phi E_3}_{M_\phi} \right)\mathbf{z} \geq 0,
\end{equation}
for some $\tau_1,\tau_2 \geq 0$. Thus, if the inequality,
$$
\mathbf{z}^\top \left( \tau_1  M_{\State_\timeid}+ \tau_2  M_{\action_\timeid}  + M_\phi \right)\mathbf{z}- \begin{bmatrix}r_{\timeid+1}\\1\end{bmatrix}^\top \begin{bmatrix} -\Omega & 0  \\ 0 &  1  \end{bmatrix} \begin{bmatrix}r_{\timeid+1}\\1\end{bmatrix} \leq 0
$$
holds, then the constraint \eqref{eq:4_1} is satisfied. This constraint can be reformulated as,
$$
\mathbf{z}^\top \!\left( \! \tau_1  M_{\State_\timeid}+ \tau_2  M_{\action_\timeid}  + M_\phi- \underline{\begin{bmatrix}C & b_\mathcal{L}\\ 0 & 1\end{bmatrix}^\top \!\!\begin{bmatrix} -\Omega & 0  \\ 0 &  1  \end{bmatrix} \begin{bmatrix}C & b_\mathcal{L}\\ 0 & 1\end{bmatrix}}_{M_{out}}\!\!\right)\mathbf{z}\! \leq 0
$$
thus applying the assumption, $ \tau_1  M_{\State_\timeid}+ \tau_2  M_{\action_\timeid}  + M_\phi-M_{out} \leq 0$, is sufficient but not necessary to claim \eqref{eq:4_1} is satisfied. Therefore, the convex optimization:
\begin{equation}\label{eq:fazlyab}
\left\{\begin{aligned}
&\min_{M_\phi, \tau_1, \tau_2}  -\mathbf{Logdet}(\Omega)\\
\ \ &\mathbf{s.t.}\  \tau_1  M_{\State_\timeid}+ \tau_2  M_{\action_\timeid}  + M_\phi-M_{out} \leq 0
\end{aligned}\right.
\end{equation}
presents the suboptimal tightest ellipsoidal upper-bound that is centered on the origin over the residual reachset $\mathcal{R}_{\timeid+1}$.
\vspace{-1mm}
% \section*{Appendix F. Linear example}
% \vspace{-1mm}
% Here the stochastic dynamics for ego car is presented for both training and deployment environments. The training environment is a stochastic linear dynamics as follows:
% $$
% \begin{aligned}
% \mathbf{x}_{t+1}&=\begin{bmatrix} 1 & 0.1 &0.0047\\0 & 1 & 0.0906 \\ 0& 0& 0.8187\end{bmatrix} \mathbf{x}_t + \begin{bmatrix} 0.003 \\ 0.0094\\ 0.1813 \end{bmatrix} u_t + \nu_t,\\
% \nu_t &\sim \gaussian\left(\begin{bmatrix}0 & 0 & 0.2 \end{bmatrix}^\top, \exp(-8)I_3 \right)
% \end{aligned}
% $$
% The deployment environment is also a stochastic linear dynamics as follows:
% $$
% \begin{aligned}
% \mathbf{x}_{t+1}&=\begin{bmatrix} 1 & 0.1 &0.0046\\0 & 1 & 0.0885 \\ 0& 0& 0.7788\end{bmatrix} \mathbf{x}_t + \begin{bmatrix} 0.004 \\ 0.0115\\ 0.2212 \end{bmatrix} u_t + \eta_t,\\
% \eta_t &\sim \gaussian\left(\vec{0}, \exp(-8)I_3 \right)
% \end{aligned}
% $$

\end{document}